%% file: main.tex
\documentclass{article}

\PassOptionsToPackage{numbers, compress}{natbib}
\usepackage[preprint]{neurips_2023}

\usepackage[utf8]{inputenc} 
\usepackage[T1]{fontenc}    
\usepackage{hyperref}       
\usepackage{url}            
\usepackage{booktabs}       
\usepackage{amsfonts}       
\usepackage{nicefrac}       
\usepackage{microtype}      
\usepackage{xcolor}         
\usepackage{amsmath}
\usepackage{amssymb}
\usepackage{mathtools}
\usepackage{amsthm}
\usepackage{multirow}
\usepackage{wrapfig}
\usepackage[capitalize,noabbrev]{cleveref}
\usepackage{graphicx}
\usepackage{subfigure} 
\usepackage{float}
\usepackage{algorithm} 
\usepackage{algorithmic}
\usepackage{textcomp} 
\usepackage{listings} 
\usepackage{enumitem}
\usepackage{utfsym}
\usepackage{times}
\usepackage{latexsym}
\usepackage{fancyvrb}
\usepackage{CJKutf8}

\newcommand{\commentout}[1]{}

\title{TouchTTS: An Embarrassingly Simple TTS Framework that Everyone Can Touch}

\author{
  Technical Report \thanks{
    Authors: Xingchen Song\textsuperscript{1,2}, Mengtao Xing\textsuperscript{1,2}, Changwei Ma\textsuperscript{1,2}, Shengqiang Li\textsuperscript{1,2}, Di Wu\textsuperscript{1,2}, Binbin Zhang\textsuperscript{1,2}, Fuping Pan\textsuperscript{1}, Dinghao Zhou\textsuperscript{2}, Yuekai Zhang\textsuperscript{2}, Shun Lei\textsuperscript{3}, Zhendong Peng\textsuperscript{2}, Zhiyong Wu\textsuperscript{3}.
    Corresponding author: Xingchen Song (sxc19@tsinghua.org.cn), Binbin Zhang (binbin.zhang@gua.com)
  } \\
  GUA Speech Team\textsuperscript{1},\quad WeNet Community\textsuperscript{2},\quad Tsinghua University\textsuperscript{3}
}

\begin{document}

\maketitle

\input{section/abstract}

\input{section/introduction}

\input{section/relatedwork}

\input{section/sec1.datapipe}

\input{section/sec2.architecture}

\input{section/sec3.unified}

\input{section/experiments}

\input{section/futurework}

\bibliographystyle{unsrt}
\bibliography{refs}

\end{document}

%% file: section/abstract.tex
\begin{abstract}
\label{sec:abs}
It is well known that LLM-based systems are data-hungry, requiring substantially more training data compared to traditional approaches.
Recent LLM-based TTS works typically employ complex data processing pipelines and elaborate filtering strategies to obtain high-quality training data.
These sophisticated pipelines require excellent models at each stage (e.g., speech denoising, speech enhancement, speaker diarization, and punctuation models), which themselves demand high-quality training data and are rarely open-sourced.
Even with state-of-the-art models, issues persist, such as incomplete background noise removal and misalignment between punctuation and actual speech pauses.
Moreover, the stringent filtering strategies often retain only 10-30\% of the original data, significantly increasing data acquisition costs and impeding data scaling efforts.
In this work, we leverage a noise-robust audio tokenizer (S3Tokenizer \cite{du2024cosyvoice}) to design a simplified yet effective TTS data processing pipeline that maintains data quality while substantially reducing data acquisition costs, achieving a data retention rate of over 50\% for the first time.
Beyond data scaling challenges, LLM-based TTS systems also incur higher deployment costs compared to conventional approaches. Current systems typically use LLMs solely for text-to-token generation, while requiring separate models (e.g., flow matching models \cite{mehta2024matcha}) for token-to-waveform generation, which cannot be directly executed by LLM inference engines, further complicating deployment.
To address these challenges, we eliminate redundant modules in both LLM and flow components, replacing the flow model backbone with an LLM architecture (e.g., Qwen2ForCausalLM \cite{yang2024qwen2}) that can be directly executed by inference engines like TensorRT \cite{TensorRT} and vLLM \cite{kwon2023efficient}. Building upon this simplified flow backbone, we propose a unified architecture for both streaming and non-streaming inference, significantly reducing deployment costs.
Finally, we explore the feasibility of unifying TTS and ASR tasks using the same data for training, thanks to the simplified data processing pipeline and the S3Tokenizer \cite{du2024cosyvoice} that reduces the quality requirements for TTS training data.
\end{abstract}

%% file: section/introduction.tex
\section{Introduction}
\label{sec:intro}

Text-to-Speech (TTS) technology has made significant progress in recent years, evolving from mechanical speech to highly natural voice output.
This advancement has been accelerated by Large Language Models (LLMs), which enable natural speech synthesis and zero-shot voice generation without speaker-specific training data \cite{dong2022survey}.
As LLM-based TTS systems develop rapidly, new challenges emerge in \textbf{data scaling} and \textbf{deployment efficiency}.

Existing industry-level LLM-based TTS systems or large-scale TTS datasets, such as CosyVoice \cite{du2024cosyvoice}, FishSpeech \cite{liao2024fish}, FireRedTTS \cite{guo2024fireredtts}, XimalayaTTS \cite{chen2024takin}, WenetSpeech4TTS \cite{ma2024wenetspeech4tts}, Emilia \cite{he2024emilia} and AutoPrep \cite{yu2024autoprep}, typically build complex data processing pipelines to convert large amounts of raw audio into high-quality TTS training data with rich annotations and diverse content and timbres.
After such complex processing, only 10\% to 30\% of the original data remains usable for training, turning data scaling into a \textbf{``money game''}.
Moreover, even with fine-grained and careful processing, it's difficult to guarantee there won't be badcases at each stage.
For instance, the denoising stage may fail to completely remove noise or background music, speaker segmentation may fail to cleanly extract target speaker's voice, or the punctuation model may produce punctuation marks that don't match the actual pauses in speech, as these punctuations are typically predicted based solely on text.
Given that we cannot guarantee perfect data quality after processing, effectively utilizing \textbf{``dirty data''} becomes key to overcoming the data scaling challenge without substantial money power.

Furthermore, current LLM-based TTS systems typically require cascading a token-to-waveform model (such as flow matching model \cite{mehta2024matcha}) to generate waveform. These cascaded models often inherit the U-Net structure from image models \cite{lipman2022flow}. This special structure not only prevents reuse of LLM inference engines, thereby increasing deployment costs, but also makes streaming inference more difficult due to its greater complexity compared to LLM structures.

To address these challenges in data scaling and model deployment faced by current LLM-based TTS systems, we propose a minimalist TTS data processing pipeline and design a deployment-oriented TTS architecture based on it.

The key contributions of our work are summarized as follows:
\begin{itemize}
    \item \textbf{\textit{Simplified Data Processing Pipeline}}: thanks to the robustness of S3Tokenizer\cite{du2024cosyvoice} against dirty data, we provide a highly simplified data processing pipeline that removes noise reduction, speech enhancement, speaker diarization, and punctuation modules.
    We then employ a simple copilot-ASR cross-validation strategy to ensure data quantity and quality, achieving a data retention rate of over 50\% and verifying data effectiveness at the million-hour level.
    \item \textbf{\textit{Simplified TTS Architecture}}: we propose a deployment-oriented architecture, including:
    \begin{itemize}
      \item Simplified frontend: we use character for Chinese and bpe for English, which reduces insertion and deletion errors compared to pure byte bpe.
      \item Qwen-based backbone for both LLM and flow: we remove the text encoder in LLM and replace the common u-net structure in flow with Qwen, making the whole model seamlessly deployable on TensorRT \cite{TensorRT} and vLLM \cite{kwon2023efficient}.
      \item Simplified and unified streaming and non-streaming strategy: thanks to the WeNet-style chunk-based inference and the aforementioned Qwen-based flow backbone, we can use a unified flow model to handle both streaming and non-streaming scenarios.
    \end{itemize}
    \item \textbf{\textit{Unified TTS and ASR}}: based on all the above simplifications, especially the data processing pipeline that significantly improves data retention rate and the S3Tokenizer \cite{du2024cosyvoice} that reduces the quality requirements for TTS training data, we explore the feasibility of unifying TTS and ASR using the same data for training.
\end{itemize}

%% file: section/relatedwork.tex
\section{Related Work}

The most related work to this paper is CosyVoice \cite{du2024cosyvoice}, which consists of an LLM for text-to-token generation and a conditional flow matching model \cite{mehta2024matcha,lipman2022flow} for token-to-speech synthesis.
Our work extends CosyVoice by simplifying the data processing pipeline and the model architecture, thereby reducing data acquisition and deployment costs, and exploring the feasibility of unified streaming and non-streaming inference, the ability of unifying TTS and ASR tasks using the same data for training.
In subsequent sections, we will detail our simplifications.

%% file: section/sec1.datapipe.tex
\section{Simplified Data Processing Pipeline}

\begin{figure}[htbp]
    \centering
    \includegraphics[width=1.0\textwidth]{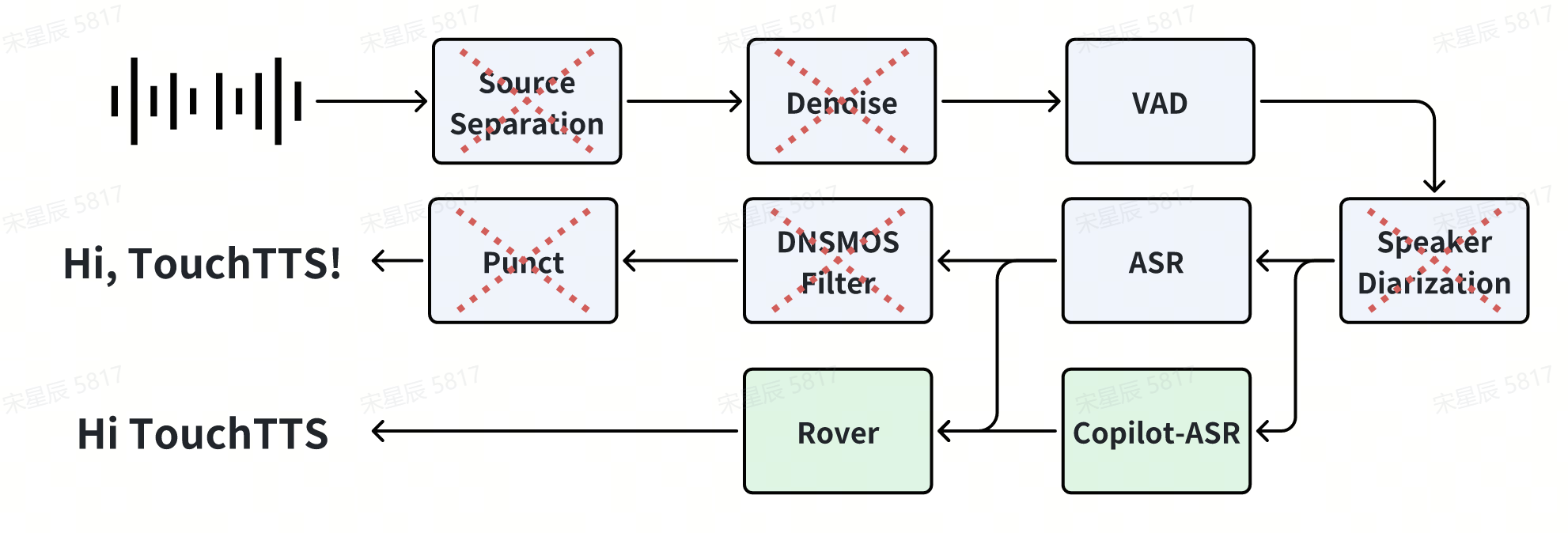}
    \caption{Overview of the Simplified Data Processing Pipeline. The blue blocks represent the common data processing pipeline of the current LLM-based TTS systems \cite{du2024cosyvoice,liao2024fish,guo2024fireredtts,chen2024takin,ma2024wenetspeech4tts,he2024emilia,yu2024autoprep}, and the red cross represents the parts removed from our pipeline. The green blocks represent the parts added to our pipeline.}
    \label{fig:pipeline}
\end{figure}

As shown in Figure \ref{fig:pipeline}, we propose a streamlined data processing pipeline for converting large-scale raw speech data into TTS training data. This pipeline consists of three primary modules: VAD, ASR, and Copilot-ASR with Rover. We first utilize VAD to segment audio into 2-30 second clips, then employ an ASR model for text transcription, followed by a secondary transcription using the Copilot-ASR model. Finally, we filter out data with significant discrepancies based on the consistency between the two transcriptions to obtain the final annotation results.

Our guiding principle is \textbf{``keeping the data pipeline as simple as possible to minimize errors and is thus more general for scaling''}. As discussed in Section \ref{sec:intro}, we cannot guarantee the reliability of each module in traditional pipelines. For instance, denoising models cannot completely eliminate background noise, speaker diarization cannot perfectly distinguish overlapping speakers, and punctuation models cannot ensure precise pause alignment. These limitations collectively reduce the overall data retention rate.

Another limitation of previous data processing pipelines is their TTS-specific customization, which alters the original data distribution, potentially making the data less suitable for tasks such as speech recognition and end-to-end spoken dialogue. For example, while denoising modules remove background sounds to benefit TTS tasks, ASR models actually achieve better robustness when trained on data containing background noise. Similarly, speaker diarization ensures single-speaker segments beneficial for TTS but may be counterproductive for end-to-end dialogue tasks, where multi-speaker interactions within single samples are crucial, particularly during the pretraining phase to facilitate better learning of speaker interactions during subsequent SFT stages.

Overall, "less is more", our proposed pipeline eliminates modules such as source separation, denoising, speaker diarization, DNSMOS \cite{reddy2022dnsmos} filtering, and punctuation, thereby preserving the original data distribution as much as possible and improving data retention rate. The effectiveness of our simplified processing pipeline is attributed to two key factors:
\begin{itemize}
    \item \textbf{S3Tokenizer for Noise-Robust Audio Tokenization.} Early studies have shown that adding ASR loss during the training of speech tokenizers outperforms the quantization method based on k-means clustering in speech translation tasks \cite{rubenstein2023audiopalm}. CosyVoice \cite{du2024cosyvoice} further demonstrates that the speech tokenizer based on ASR loss (Supervised Semantic Speech Tokenizer, S3Tokenizer) is also effective for TTS tasks. In this paper, we further explore the potential of S3Tokenizer in data scaling. We argue that since the optimization goal of ASR loss is to extract semantic information from speech and ignore irrelevant background noise, speaker information, etc., the S3Tokenizer with ASR loss implicitly learns the ability to denoise and disentangle speakers, making it more robust to dirty data and thus reducing the quality requirements for TTS training data. Based on this discovery, we simplify our data processing pipeline.
    \item \textbf{Copilot-ASR Cross-Validation for Data Quality Assurance.} We find that a single ASR model may make errors, but the probability of two ASR models making errors is significantly lower. Moreover, data of low quality tends to exhibit significant discrepancies across different ASR models \cite{zhang2024conformer,ramirez2024anatomy}. Therefore, we filter out data with significant discrepancies based on the consistency between the two transcriptions to improve data quality. We hypothesize that this cross-validation approach implicitly assumes the functions of traditional SNR filtering and DNSMOS \cite{reddy2022dnsmos} filtering. Specifically, we implement the Rover function by comparing the Word Error Rate (WER) and Phoneme Error Rate (PER) between two transcriptions and filter out data with WER greater than 10 and PER greater than 5. We use Whisper \cite{whisper} and Paraformer \cite{gao2022paraformer} as ASR and Copilot-ASR models.
\end{itemize}

\begin{figure}[htbp]
    \centering
    \includegraphics[width=1.0\textwidth]{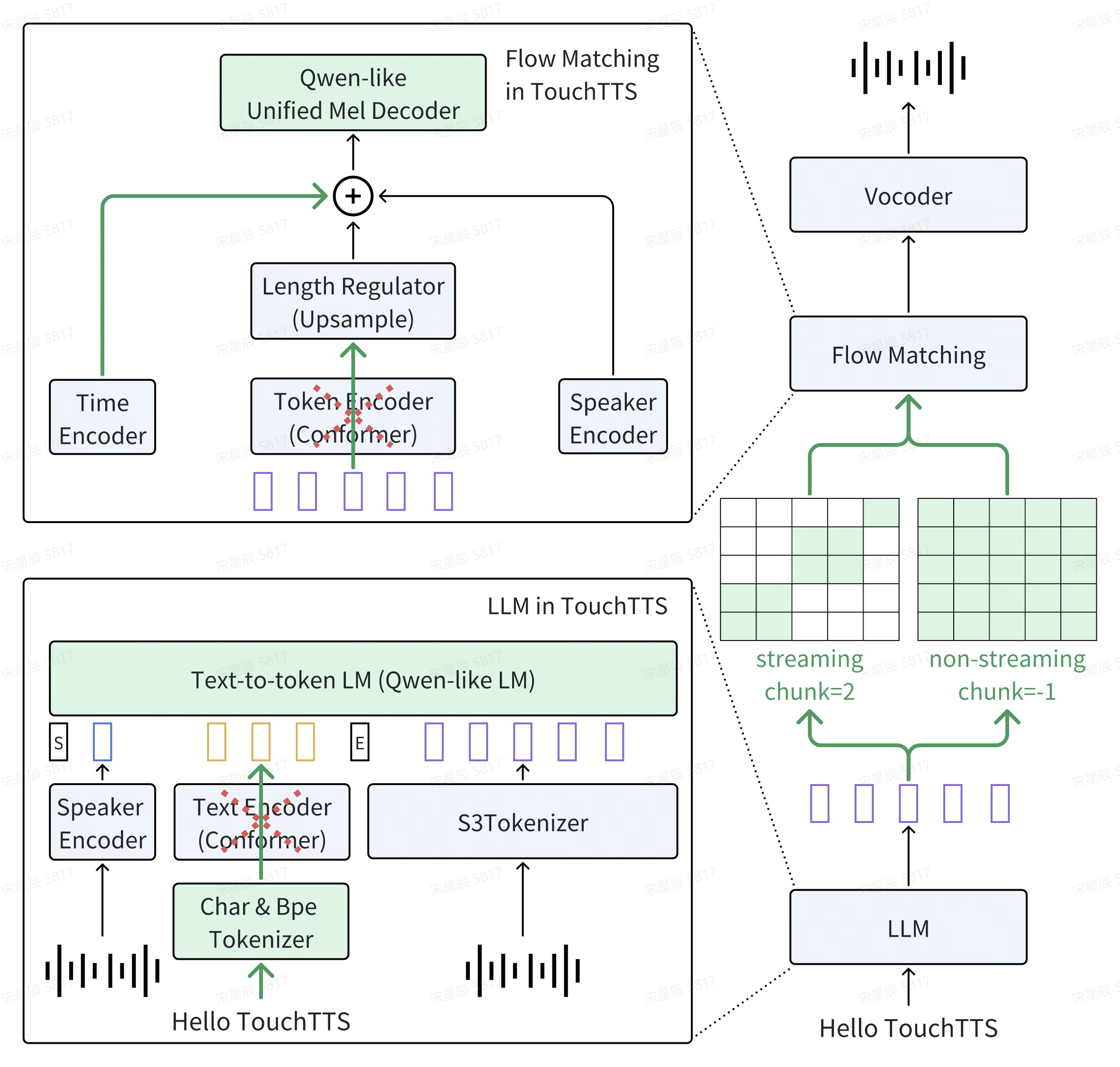}
    \caption{Overview of the Simplified TTS Architecture. The blue blocks represent the common LLM-based TTS architecture, and the red cross represent the parts removed from our architecture. The green blocks and green lines represent the parts that we added or simplified. "S" denotes start of the sentence and "E" denotes end of the sentence. Note that in the streaming mode, the chunk size can be arbitrary, here we assume it is 2 for convenience.}
    \label{fig:architecture}
\end{figure}

Finally, we process 1260k hours of raw data into 650k hours of training data, achieving a data retention rate of 51.6\%. Notably, even with the same data processing pipeline, the data retention rate varies significantly across different domains. For instance, audiobooks, with relatively consistent recording environments and less background noise, generally have a higher retention rate than outdoor live broadcast data. The 51.6\% retention rate is an average across all domains.

In addition, we collected 150k hours of open-source data (mostly ASR data) and supplemented it with some internal data (also mostly ASR data) to form our final one million hours of training data (85\% Chinese and 15\% English).

%% file: section/sec2.architecture.tex
\section{Simplified TTS Architecture}


As shown in Figure \ref{fig:architecture}, we simplify the LLM-based TTS architecture, mainly in the following three aspects.

We use character-based units for Chinese text tokenization and keep the English text tokenization as BPE. We will show in Section \ref{sec:experiment} that this improvement can help reduce the number of insertion and deletion errors in Chinese TTS.

We use Qwen2ForCausalLM~/~Qwen2MoeForCausalLM \cite{yang2024qwen2} as our LLM backbone and Qwen2ForCausalLM \cite{yang2024qwen2} as our flow backbone. At the same time, we remove the Text Encoder \cite{du2024cosyvoice,chen2024takin} in the LLM part and the Token Encoder \cite{du2024cosyvoice,guo2024fireredtts} in the flow part to further simplify the model structure.

Since the flow part no longer uses the U-net structure \cite{du2024cosyvoice,guo2024fireredtts,mehta2024matcha} and does not have the cross-layer skip connections, we can easily unify the streaming and non-streaming inference by controlling the attention mask of the input, which is inspired by the related work in the ASR field \cite{zhang2022wenet}, allowing us to implement a flow model that supports both streaming and non-streaming modes through WeNet-style chunk-based inference \cite{zhang2022wenet}.

\subsection{Character-based Unit for Simple yet Efficient Chinese Text Tokenization}
\label{sec:frontend}

Traditional TTS architectures, including VALL-E \cite{wang2023neural}, VITS \cite{kim2021conditional}, and FastSpeech \cite{FastSpeech}, typically rely on grapheme-to-phoneme (G2P) conversion to transform text into phonetic representations.
While this approach has proven effective, it faces significant challenges when handling context-dependent polyphonic words and cross-lingual generalization due to the complexity of phonetic rules.
As the demand for LLM-based large-scale TTS systems continues to grow, the limitations of G2P-based approaches have become increasingly apparent.
The heavy reliance on language-specific phonetic rules and lexicons significantly constrains system scalability and introduces substantial maintenance overhead.
Recent research, such as FishSpech \cite{liao2024fish}, CosyVoice \cite{du2024cosyvoice}, FireRedTTS \cite{guo2024fireredtts}, have made promising strides by exploring the use of bpe-based tokenizers from Large Language Models (LLMs) for direct linguistic feature extraction, effectively eliminating the need for explicit G2P conversion.

In this work, we find that character-based units can be more simple and efficient for Chinese text tokenization. This is due to the following reasons:
\begin{itemize}
    \item The pronunciation rules of Chinese are that one character corresponds to one complete pronunciation, so the number of tokens obtained by character modeling is exactly the same as the number of pronunciations, which explicitly provides a strong bias to the model. In contrast, when using BPE to model Chinese, one Chinese character may be tokenized into more than one or less than one token, resulting in an inconsistent number of tokens and pronunciations for the same text, which makes the model lose the guiding prior information and is more prone to insertion and deletion errors.
    \item There are many homophones in Chinese, which have the same pronunciation in phonetic terms but are completely different in character terms. Therefore, using character-based units can better capture the differences between homophones than using G2P methods.
    \item There are many polyphonic characters in Chinese, which may have different pronunciations in different contexts. Compared to using G2P methods, using character-based units can better handle the pronunciation of polyphonic characters in different contexts.
\end{itemize}

\subsection{Pure Qwen Backbone for both LLM and Flow}





Our design principle is to \textbf{leverage the LLM ecosystem as much as possible}, including training libraries (such as transformers \cite{wolf-etal-2020-transformers}) and inference libraries (such as TensorRT \cite{TensorRT}, vLLM \cite{kwon2023efficient}). Therefore, we mainly make architecture improvements in two aspects: replacing the backbones of LLM and Flow with standard transformer structures, and removing unnecessary modules in LLM and Flow.

Specifically, as shown in Figure \ref{fig:architecture}, we use Qwen \cite{yang2024qwen2} as the backbone of LLM and Flow. It is worth noting that the backbone can be any LLM structure supported by TensorRT, vLLM, etc., here we only conduct experiments on Qwen structure for simplicity.
It is a straightforward idea to apply the LLM structure to the text-to-token LM, because the text-to-token LM is essentially a sequence-to-sequence task, and LLM is designed for sequence-to-sequence tasks. However, for the Flow part, the U-net structure is widely used and has become a de facto standard in the image \cite{lipman2022flow} and speech \cite{du2024cosyvoice,guo2024fireredtts,mehta2024matcha,miao2020flow} fields.
Although the effectiveness and performance of the U-net structure have been widely verified in the flow part, its complex structure and cross-layer skip connections not only make it difficult to reuse the LLM inference ecosystem but also pose significant challenges for streaming inference.

On the contrary, using the transformer structure in the Flow can not only better reuse the LLM inference ecosystem but also easily achieve the unification of streaming and non-streaming through controlling the attention mask. However, whether the synthesis result can match that of the U-net structure is an open question.
In this paper, we verify through experiments that using the transformer structure in the Flow is effective and efficient, and can achieve the unification of streaming and non-streaming.

Another factor that hinders the efficient deployment of LLM-based TTS systems is the redundant modules in the architecture design, such as the common Text Encoder (Conformer) \cite{du2024cosyvoice,chen2024takin} in the LLM part and the common Token Encoder (Conformer) \cite{du2024cosyvoice,guo2024fireredtts} in the Flow part. These modules are usually not supported by inference libraries such as TensorRT \cite{TensorRT}, vLLM \cite{kwon2023efficient}, etc., requiring additional deployment, which not only increases the complexity of the system but also reduces the inference efficiency.
At the same time, we believe that when the parameter size of the backbone is large enough, the complexity of other modules in the model has a negligible impact on the final effect, so we simplify the model structure as much as possible to better reuse the LLM inference ecosystem.

\subsection{Chunk-based Unified Streaming and Non-Streaming Inference}

\begin{figure}[!htp]
    \centering
    \includegraphics[width=0.8\textwidth]{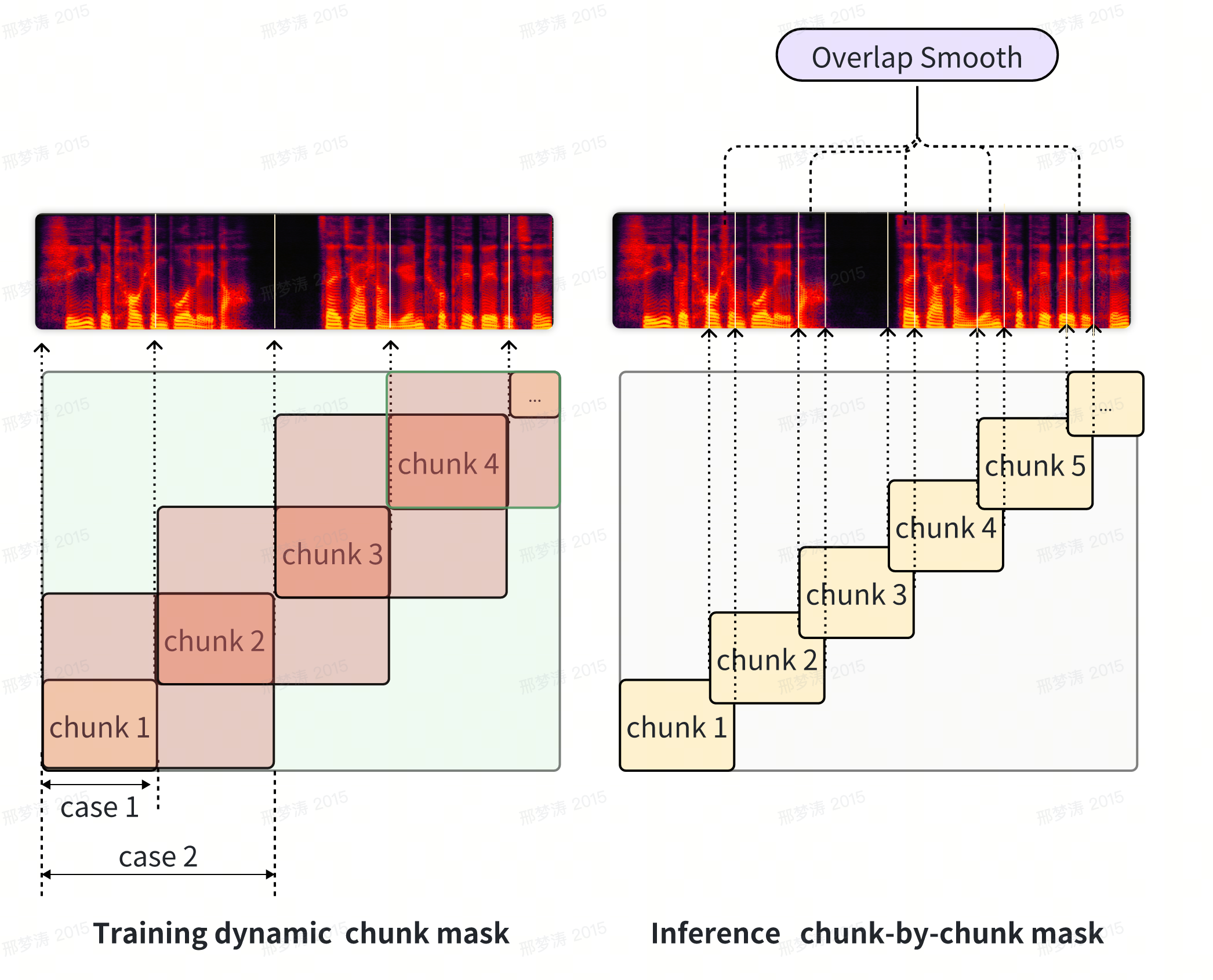}
    \caption{Illustration of the training and inference pipeline for unified streaming and non-streaming synthesis. As shown in the figure, during training (left), after selecting dynamic chunk lengths, Case 1 represents chunks without overlap, while Case 2 represents chunks with one chunk length of historical context overlap. The receptive field of the current chunk is masked, with chunk length and historical context length varying dynamically during training. During inference (right), the current chunk's receptive field is set to full coverage, and shorter token overlaps are used between chunks to smooth mel-spectrogram boundaries.}
    \label{fig:flow_explain}
\end{figure}


To simultaneously support both streaming and non-streaming inference modes, inspired by the dynamic chunk training approach in WeNet\cite{zhang2022wenet}, we concatenate conditional information such as speaker embeddings and time embeddings into the input features of our flow model, which are then uniformly fed into the flow module. Our flow module employs the Qwen model as its backbone without an additional token encoder. During training, we replace the causal mask with a dynamic chunk mask, which ensures that the receptive field is shared within the current chunk segment while maintaining independence from future information. The historical context length for each chunk segment varies dynamically based on the chunk length, allowing flexibility in utilizing or disregarding historical information. This approach enables the model to adapt to both chunk-by-chunk inference scenarios and those requiring historical context, thereby bridging the receptive field discrepancy between training and streaming inference. In our experiments, we set the minimum chunk length to 0.5 seconds and dynamically expand it during training, maintaining sentence-level receptive fields for 50\% of the training data to simultaneously accommodate varying chunk receptive fields while ensuring training stability.


During the model inference phase, different token lengths can be configured as chunks. Once the LLM model generates tokens for the current chunk, the flow module can proceed with inference without concerns about insufficient future token information affecting the performance. Notably, when inferring tokens for the current chunk, consistency with the training phase is maintained - all information within the current chunk segment remains visible, resulting in a full receptive field mask for the current chunk. This alignment between the receptive fields during inference and training effectively bridges any discrepancies between these phases. While we adopt the same streaming inference token overlap mechanism as CosyVoice \cite{du2024cosyvoice}, our model has been trained to accommodate tokens with varying receptive field lengths and has been exposed to training data with different token overlap configurations, including scenarios without overlap (i.e., zero historical context length). Consequently, we can employ shorter token overlaps to smooth the mel-spectrogram boundaries between different chunks during inference.


%% file: section/sec3.unified.tex
\section{Unified TTS and ASR}

\begin{figure}[htbp]
    \centering
    \includegraphics[width=0.9\textwidth]{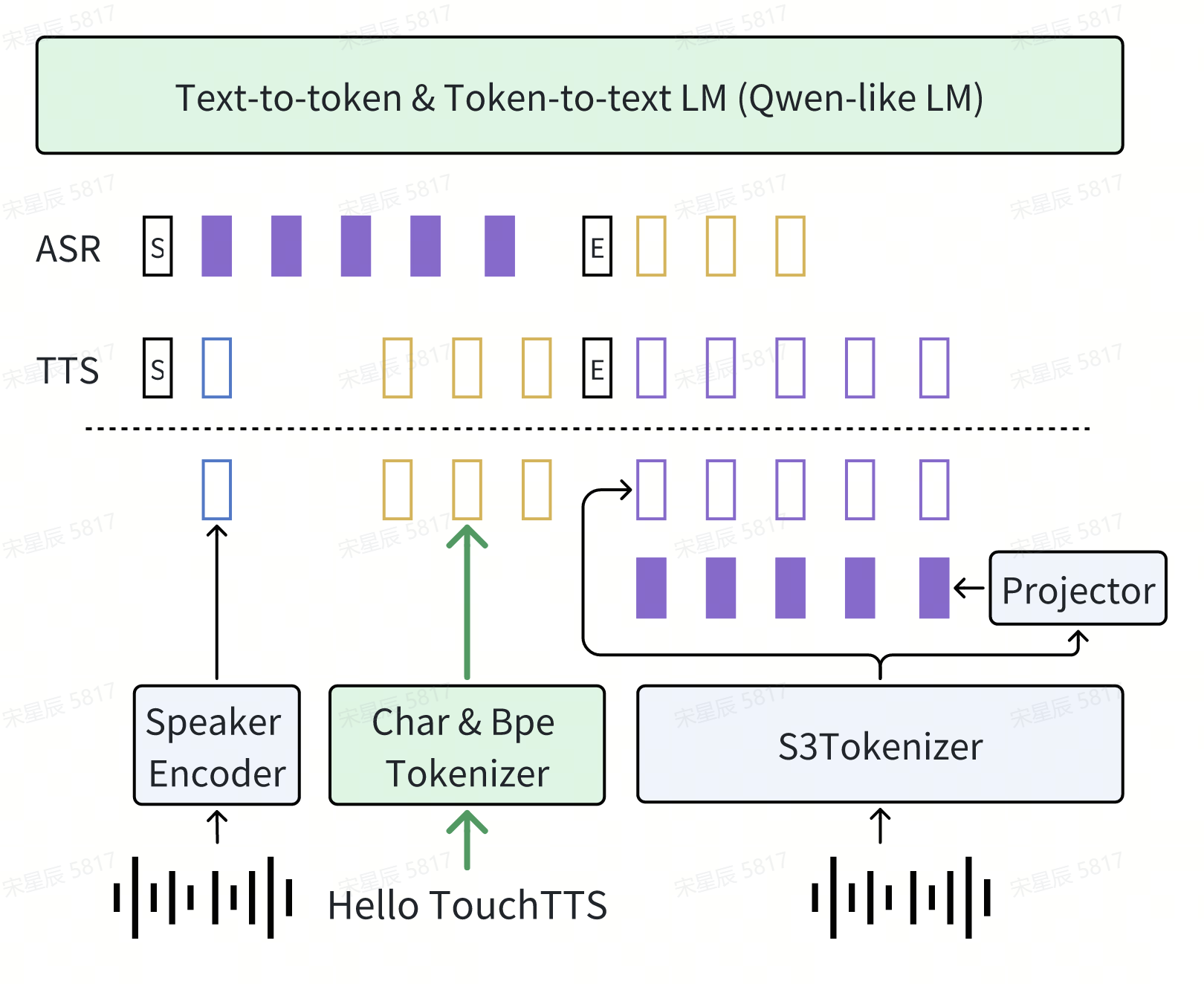}
    \caption{Overview of Unified TTS and ASR. In the unified architecture, ASR and TTS share the same LLM, but process their own inputs and outputs separately. The input of TTS is the speaker embedding (blue hollow square) and text token (yellow hollow square), and the output is the audio token (purple hollow square) encoded by S3Tokenizer. The input of ASR is the last layer hidden output of S3Tokenizer transformed by projector \cite{ma2024embarrassingly} (purple solid square), and the output is the text token.}
    \label{fig:unified}
\end{figure}


As shown in Figure \ref{fig:unified}, we propose a unified architecture that allows an LLM to perform both Text-to-token and Token-to-text tasks. Previous work (e.g., LauraGPT \cite{chen2023lauragpt}, Seed-ASR \cite{bai2024seed}) has shown that continuous features for audio have notable advantages over discrete units on speech recognition, understanding, and audio-signal related tasks.
We extend this conclusion and verify it on our unified TTS and ASR tasks.

Although LauraGPT \cite{chen2023lauragpt} adopts a similar unified architecture, the training data for different tasks are independent of each other (the ASR data is not used in TTS, and vice versa).
In this paper, we attempt to train TTS and ASR tasks simultaneously using the same large-scale dirty data for the first time (notably, dMel \cite{bai2024dmel} only validated the effectiveness of unified ASR and TTS tasks on clean, small-scale reading-style speech data like LibriSpeech \cite{panayotov2015librispeech}, and its effectiveness on large-scale dirty data remains to be verified), and verify the effectiveness of this approach. This is mainly due to the improvement in data retention rate brought by the simplified data pipeline and the reduction in the quality requirements for TTS data brought by S3Tokenizer \cite{du2024cosyvoice}, allowing data that meets the ASR standard to be used for TTS tasks.

Another work similar to our unified architecture is GLM4-Voice \cite{zeng2024scaling}, SpiritLM \cite{nguyen2024spirit}, which unifies ASR and TTS tasks in the same architecture, but encodes the input of the ASR task as discrete tokens, which prevents it from utilizing the advantages of continuous features.
Moreover, SpiritLM uses \textbf{word-level speech-text interleaving} at the input, which requires an additional alignment model to provide word-level timestamps.
Conversely, our approach uses \textbf{sentence-level speech-text interleaving} at the input without the need for an additional alignment model, simplifying the data preparation process.

%% file: section/experiments.tex
\section{Experiments}
\label{sec:experiment}


We mainly conduct our experiments on Seed-Eval \cite{anastassiou2024seed}. Seed-Eval consists of three parts: \textit{test-zh}, \textit{test-en}, \textit{test-hard}. Among them, \textit{test-zh} contains 2000 Chinese samples extracted from the DiDiSpeech dataset \cite{guo2021didispeech}, \textit{test-en} contains 1000 English samples extracted from the Common Voice dataset \cite{ardila2019common}, and \textit{test-hard} contains 400 sentences with especially challenging patterns for autoregressive models such as word repetitions, tongue twisters, and so on.

We provide some samples from the three test sets below to help readers understand the difficulty of the different test sets.

\begin{itemize}
    \item \textit{test-en}: I heard the land where the hobbits live\textcolor{red}{\textbf{,}} the Shire\textcolor{red}{\textbf{,}} has actually been filmed in New Zealand\textcolor{red}{\textbf{.}}
    \item \textit{test-zh}: \begin{CJK}{UTF8}{gbsn}天黑了\textcolor{red}{\textbf{，}}路灯亮起\textcolor{red}{\textbf{，}}雪雾扬起\textcolor{red}{\textbf{，}}寒冷像潮水一样涌来\textcolor{red}{\textbf{。}}\end{CJK}
    \item \textit{test-hard} (word repetitions): \begin{CJK}{UTF8}{gbsn}被称赞了了了了了了了真开心开心开心开心开心啊啊啊啊啊啊\textcolor{red}{\textbf{，}}我我我我我有那么帅该多多多多多多好好好好好好好啊啊啊啊啊啊啊\textcolor{red}{。}\end{CJK}
    \item \textit{test-hard} (tongue twisters): \begin{CJK}{UTF8}{gbsn}夫妇做豆腐\textcolor{red}{\textbf{，}}夫妇卖豆腐\textcolor{red}{\textbf{。}}夫富妇也富\textcolor{red}{\textbf{，}}富妇也富夫\textcolor{red}{\textbf{。}}\end{CJK}
\end{itemize}

It is important to note that our model did not use any punctuation during the training process. Therefore, the actual input for our model is shown below, which has had the punctuation removed. Generally, researchers believe that training speech synthesis models on text inputs without punctuation as explicit prosodic supervision is difficult. However, our experiments show that, with a large amount of data, autoregressive models can still produce audio prosody and pauses that sound natural, even without explicit punctuation to guide the synthesis.

\begin{itemize}
    \item \textit{test-en}: I heard the land where the hobbits live the Shire has actually been filmed in New Zealand
    \item \textit{test-zh}: \begin{CJK}{UTF8}{gbsn}天黑了路灯亮起雪雾扬起寒冷像潮水一样涌来\end{CJK}
    \item \textit{test-hard} (word repetitions): \begin{CJK}{UTF8}{gbsn}被称赞了了了了了了了真开心开心开心开心开心啊啊啊啊啊啊我我我我我有那么帅该多多多多多多好好好好好好好啊啊啊啊啊啊啊\end{CJK}
    \item \textit{test-hard} (tongue twisters): \begin{CJK}{UTF8}{gbsn}夫妇做豆腐夫妇卖豆腐夫富妇也富富妇也富夫\end{CJK}
\end{itemize}

As for evaluation metrics, we use Paraformer-en \cite{gao2022paraformer} to transcribe English and Paraformer-zh \cite{gao2022paraformer} to transcribe Chinese and calculate Phoneme Error Rate (PER) respectively. We use pypinyin\footnote{https://github.com/mozillazg/python-pinyin} for Chinese G2P and g2p-en\footnote{https://github.com/Kyubyong/g2p} for English G2P.

Here, we do not use Whisper-large-v3 \cite{whisper} for English transcription because Whisper performs Inverse Text Normalization (ITN) internally, which may affect the results.
Besides, we no longer use WER as the evaluation metric chosen by most papers \cite{du2024cosyvoice,liao2024fish,anastassiou2024seed,chen2024f5,wang2024maskgct}, because we find that WER does not accurately reflect the content intelligibility and pronunciation correctness of synthesized speech.
For example, the recognition results of ASR systems often have homophone errors. In extreme cases, if the pronunciation of synthesized speech is completely correct in human evaluation, but the ASR system's transcription results are all homophones, then the WER will be 100\%, while the PER will be 0\%.
Therefore, we believe that PER is a better metric for reflecting the content intelligibility and pronunciation correctness of synthesized speech than WER.

For each test set, we conduct tests using five different seeds, ranging from 2024 to 2028. For the PER metric, we aggregate the insertion, deletion, and substitution errors across the five groups and then compute the average PER.

For the Speaker Similarity (SIM) metric, we use wespeaker \cite{wang2023wespeaker} to extract speaker embedding and simply take the average of the five cosine similarity results.

\subsection{Frontend comparison: bpe vs character-based modeling}
\label{sec:frontendexp}


In this section, to control the experimental variables, we reuse the S3Tokenizer\footnote{https://github.com/xingchensong/S3Tokenizer} and all flow modules from CosyVoice \cite{du2024cosyvoice}, which do not participate in the training process. We only remove the text encoder and replace the backbone of the LLM with Qwen2ForCausalLM (around 500M parameters, denoted as TouchLLM-0.5B) and train it on 1M hours for just one epoch to compare the effects of different modeling methods.

\begin{table}[htp]
  \caption{Comparison between different modeling units.}
  \label{FrontendCompare}
  \centering
  \begin{tabular}{cc|c|cccc}
    \toprule
    Unit for zh & Unit for en  & TestSet & PER(\%)  & Sub & Del & Ins  \\
    \midrule
    bpe & bpe  & \textit{test-zh} & 1.275 & \textbf{1971} & 574 & 150 \\
    char & bpe & \textit{test-zh} & \textbf{1.124} & 2068 & \textbf{222} & \textbf{86} \\
    \midrule
    bpe & bpe & \textit{test-hard} & 8.833 & 3088 & 4820 & 1399  \\
    char & bpe & \textit{test-hard} & \textbf{7.96} & \textbf{2982} & \textbf{4580} & \textbf{830} \\
    \bottomrule
  \end{tabular}
\end{table}

In Table \ref{FrontendCompare}, the results show that by changing the Chinese modeling unit from bpe to char, a significant reduction in total deletion and insertion errors can be observed across different Chinese test sets. This improvement is attributed to the strong bias in the number of pronunciations provided by character modeling, as mentioned in the section \ref{sec:frontend}.

\subsection{Model comparison with baselines}


In this section, we use the same TouchLLM-0.5B model as in \ref{sec:frontendexp} to compare with other TTS systems.
We know that comparing with these open-source models is unfair because the model sizes, training times, and data durations differ.
To minimize these differences as much as possible, we limit the number of training epochs.
This model is trained for only one epoch, the training cost is equivalent to training for 10 epochs using a 100,000-hour dataset (e.g., Emilia \cite{he2024emilia}).


In Table \ref{ModelCompare}, we use the NoPrompt suffix to indicate that only speaker embedding is used during inference, without using prompt text and prompt wav. The model with the ZeroShot suffix, on the other hand, indicates that prompt text and prompt wav are used during inference.
By comparing the two different decoding methods, we find that the ZeroShot method can reduce PER on simpler test sets like \textit{test-zh} and \textit{test-en}, but it significantly increases deletion errors on the more difficult test set \textit{test-hard}.
We believe this is due to the training-inference mismatch present in the ZeroShot method. For example, on one hand, \textit{test-hard} contains many artificially constructed sentences with repetitions, which are difficult to appear in training. On the other hand, the prompt text used in the ZeroShot method may be semantically incoherent with the text to be synthesized, further increasing the difficulty of model decoding.
Another conclusion we can draw from Table \ref{ModelCompare} is that the ZeroShot method can significantly enhance speaker similarity, thanks to the powerful in-context learning ability of the LLM.


Additionally, our model typically exhibits significantly more deletion errors and fewer insertion errors. The high deletion rate indicates a word skipping phenomenon when our model encounters a stack of repeating words. The low insertion rate makes it clear that our model is free of endless repetition.

\begin{table}[htp]
  \caption{Objective comparison between different models across different testsets.}
  \label{ModelCompare}
  \centering
  \begin{tabular}{c|cccc|c}
    \toprule
    Model & PER(\%) & Sub & Del & Ins & SIM \\
    \midrule
    \multicolumn{6}{c}{\textit{test-zh}} \\
    \midrule
    FireRedTTS\cite{guo2024fireredtts} & \textbf{0.51} & 648 & 357 & 63 & 0.850 \\
    CosyVoice-NoPrompt\cite{du2024cosyvoice} & 1.29  & 2367 & 256 & 113 & 0.802 \\
    CosyVoice-ZeroShot\cite{du2024cosyvoice} & 1.10  & 1950 & 283 & 101 & 0.913 \\
    TouchLLM-0.5B-NoPrompt & 1.12 & 2068 & 222 & 86 & 0.837 \\
    TouchLLM-0.5B-ZeroShot & 1.05 & 1940 & 197 & 90 & \textbf{0.915} \\
    \midrule
    \multicolumn{6}{c}{\textit{test-en}} \\
    \midrule
    FireRedTTS\cite{guo2024fireredtts} & 6.42 & 5116 & 2115 & 8132 & 0.754 \\
    CosyVoice-NoPrompt\cite{du2024cosyvoice} & 5.77 & 7976 & 3632 & 2194 & 0.755 \\
    CosyVoice-ZeroShot\cite{du2024cosyvoice} & 5.79 & 7843 & 3405 & 2601 & 0.859 \\
    TouchLLM-0.5B-NoPrompt & 4.81 & 6671 & 2697 & 2151 & 0.764\\
    TouchLLM-0.5B-ZeroShot & \textbf{4.55} & 6277 & 2581 & 2040 & \textbf{0.866} \\
    \midrule
    \multicolumn{6}{c}{\textit{test-hard}} \\
    \midrule
    FireRedTTS\cite{guo2024fireredtts} & 18.67 & 2332 & 15554 & 1783 & 0.846 \\
    CosyVoice-NoPrompt\cite{du2024cosyvoice} & 11.08 & 4195 & 2933 & 4547 & 0.821 \\
    CosyVoice-ZeroShot\cite{du2024cosyvoice} & 11.29 & 4497 & 3274 & 4125 & 0.906 \\
    TouchLLM-0.5B-NoPrompt & \textbf{7.96} & 2982 & 4580 & 830 & 0.835 \\
    TouchLLM-0.5B-ZeroShot & 9.62 & 3176 & 6016 & 944 & \textbf{0.910} \\
    \bottomrule
  \end{tabular}
\end{table}

\subsection{Analysis of streaming inference results}
In this section, we conduct a comparative analysis between streaming and non-streaming performance of the Qwen flow model trained with dynamic chunk strategy (denoted as TouchFlow). To investigate the impact of model size on flow performance, we trained two models with different parameter configurations: TouchFlow-170M and TouchFlow-50M, containing 170M and 50M parameters respectively. Both models were trained on 50,000 hours of English and Chinese data from the Emilia dataset \cite{he2024emilia}.

For fair comparison, both models were trained with identical hyperparameter settings, including a learning rate of 1e-3 and a warm-up initialization strategy over 5,000 steps. The models were trained for 2 epochs (600k steps) using the same dynamic chunk configuration.

To ensure experimental reproducibility, we employed CosyVoice's \cite{du2024cosyvoice} 25Hz LLM model as the speech token model, maintaining consistent sampling results across experiments by using identical random seeds. The comparative analysis of our two flow model variants was conducted under these controlled conditions, with both models utilizing 5 flow iterations during inference.

As shown in Table \ref{StreamCompare}, we employ PER and SIM as objective metrics to evaluate speaker similarity and intelligibility across different streaming configurations. To demonstrate the effectiveness of our dynamic chunk strategy, we compare the performance between non-streaming and various streaming inference configurations.

The notation "TouchFlow-170M non-streaming" represents results from the TouchFlow-170M model under non-streaming inference. For streaming configurations, we use a format like "TouchFlow-170M (1 2 5)", which indicates the initial chunk size is 1s (25 speech tokens), the subsequent chunk size is 2s (50 speech tokens) and 5 speech tokens between adjacent chunks for smooth.

The experimental results reveal that in non-streaming scenarios, TouchFlow-170M marginally outperforms TouchFlow-50M, though the performance gap remains relatively small. Notably, across various streaming configurations, the impact on speaker similarity and PER metrics is minimal, with performance largely consistent with non-streaming results.

This consistency demonstrates the robustness of our flow model trained with dynamic chunk strategy in effectively bridging the receptive field discrepancy between training and inference phases.

\begin{table}[htp]
  \caption{Objective comparison between streaming and non-streaming inference results.}
  \label{StreamCompare}
  \centering
  \begin{tabular}{c|cc|cc}
    \toprule
    & \multicolumn{2}{c|}{\textit{test-en}} & \multicolumn{2}{c}{\textit{test-zh}} \\
    \midrule
    Stream Config & PER(\%) & SIM & PER(\%) & SIM \\
    \midrule
    TouchFlow-170M non-streaming   &5.39  &0.828 &1.03	&0.880 \\
    \midrule
    TouchFlow-170M (1 2 5)         &5.45  &0.827 &0.94	&0.879 \\
    \midrule
    TouchFlow-170M (1 2 10)        &5.71  &0.828 &0.94 &0.880  \\
    \midrule
    TouchFlow-170M (2 4 5)         & 5.21  &0.829 &1.03 &0.881 \\
    \midrule
    TouchFlow-170M (2 4 10)        & 5.51  &0.828 &0.96 & 0.880 \\
    \midrule
    \midrule
    TouchFlow-50M non-streaming   &5.49   &0.819 &1.04	&0.867 \\
    \midrule
    TouchFlow-50M (1 2 5)         & 5.76   &0.818 &1.10	& 0.866 \\
    \midrule
    TouchFlow-50M (1 2 10)        & 5.84  &0.819 &1.05 &0.866 \\
    \midrule
    TouchFlow-50M (2 4 5)         & 5.82   &0.818 &1.05 & 0.867 \\
    \midrule
    TouchFlow-50M (2 4 10)        & 5.53   &0.818 &0.94	& 0.867 \\
    \bottomrule
  \end{tabular}
\end{table}


\subsection{Inference benchmark}
\label{sec:inference}

\begin{table}[htp]
  \centering
  \caption{Performance Comparison of FP16 and FP32 on TouchLLM-0.5B.}
  \begin{tabular}{c|c|c|c|c}
    \toprule
    Operation & Input Length & Output Length & {FP16(tokens/s)} & {FP32(tokens/s)} \\
    \midrule
    \multirow {3} * {Prefill} & 128 & 1 & 23188.40 & 9460.46 \\
    & 1024 & 1 & 46334.80 & 11318.70 \\
    & 2048 & 1 & 47178.10 & 10316.90 \\
    \midrule
    \multirow {3} * {Decode} & 1 & 128 & 432.83 & 286.49 \\
    & 1 & 1024 & 404.98 & 270.91 \\
    & 1 & 2048 & 395.97 & 257.20 \\
    \bottomrule
    \end{tabular}
  \label{tab:speedllm}
\end{table}

\begin{table}[htp]
  \centering
  \caption{Performance Comparison of FP16 and FP32 on TouchFlow-170M.}
  \begin{tabular}{c|c|c}
    \toprule
    Operation & Sequence Length &  FP32 (tokens/s) \\
    \midrule
    \multirow{3}{*}{Prefill} & 128 & 25196.90 \\
    & 1024 & 30880.60 \\
    & 2048 & 27208.70 \\
    \bottomrule
  \end{tabular}
  \label{tab:speedflow}
\end{table}

We conducted performance testing using TRT-LLM on single RTX 3090. First, we performed speed tests on the two most computationally intensive modules: the LLM (TouchLLM-0.5B) and flow (TouchFlow-170M), with results shown in Table \ref{tab:speedllm} and \ref{tab:speedflow}. 

Additionally, we tested the LLM module on our internal text dataset, with text lengths ranging from 2 to 150 tokens. We using a chunk size of 25 tokens for the first chunk. The results of First Chunk Generation are presented in Table \ref{tab:firstchunkgeneration}.

\begin{table}[htp]
  \centering
  \caption{Latency Analysis for First Chunk Generation (TouchLLM-0.5B, FP16 mode).}
  \begin{tabular}{ccccccc}
    \toprule
    Configuration & Max (ms) & Min (ms) & Avg (ms) & P95 & P98 & P99 \\
    \midrule
    1 thread & 113.75 & 5.43 & 44.78 & 73.27 & 74.36 & 76.59 \\
    \bottomrule
  \end{tabular}
  \label{tab:firstchunkgeneration}
\end{table}

With the step count set to 5, the flow component (TouchFlow-170M) has a latency of approximately 50ms, and the total latency of both modules peaks at around 160ms. Since these two modules represent the most computationally intensive components, this indicates we can readily achieve first-packet latency within 200ms.

\subsection{Evaluation on unified TTS \& ASR}

\begin{figure}[htbp]
    \centering
    \includegraphics[width=0.9\textwidth]{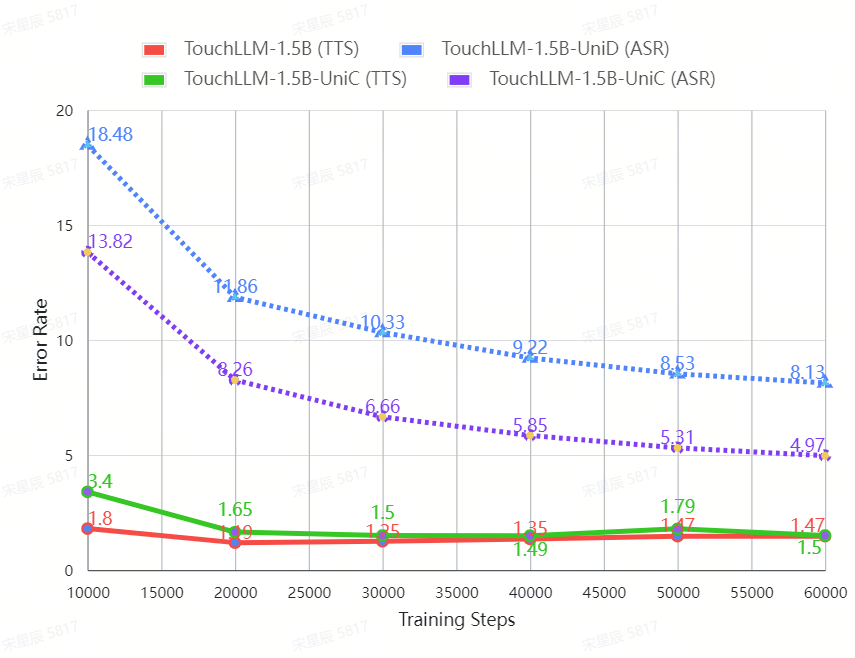}
    \caption{Comparison between single TTS model and unified TTS/ASR model. For TTS task, we calculate PER on \textit{test-zh}, for ASR task, we calculate CER on \textit{SpeechIO} \cite{SpeechIO}. For simplicity, we only count the error rate trend for the first 60k steps (approximately 0.4M hours training data), as these trends are sufficient to demonstrate the advantages and disadvantages of different methods.}
    \label{fig:unifiedexp}
\end{figure}

In this section, similar to Section \ref{sec:frontendexp}, we only train the LLM part, while replacing the backbone with Qwen2MoeForCausalLM (approximately 1.5B parameters, denoted as TouchLLM-1.5B).
For comparison, we trained two unified models with the same parameter and the same data, namely TouchLLM-1.5B-UniC and TouchLLM-1.5B-UniD, where UniC indicates ASR training using continuous features, and UniD indicates ASR training using discrete tokens.
The main purpose of this section's experiments is to answer the following two questions:
\begin{itemize}
  \item Can the unified model achieve similar performance in TTS tasks compared to independent TTS models?
  \item In the unified model, which performs better for ASR tasks: using continuous features or using discrete tokens?
\end{itemize}

Based on the results in Figure \ref{fig:unifiedexp}, we can answer the above questions:
\begin{itemize}
  \item By comparing the performance of TouchLLM-1.5B and TouchLLM-1.5B-UniC in TTS tasks, we can conclude that the unified model can perform on par with the standalone model in TTS tasks.
  \item By comparing the performance of TouchLLM-1.5B-UniC and TouchLLM-1.5B-UniD in ASR tasks, continuous features significantly outperform discrete features for understanding tasks, this conclusion is in line with LauraGPT \cite{chen2023lauragpt}.
\end{itemize}


%% file: section/futurework.tex
\section{Future Works}


Our work can be regarded as a pre-trained version of a general-purpose speech foundation model,
on top of which, we can conduct Supervised Fine-Tuning (SFT) to adapt to specific application scenarios.
Therefore, future work can be carried out in the following aspects:
\begin{itemize}
    \item Fine-grained style control (emotion, breath, break, etc.)
    \item Generation with text instruction
    \item End-to-end speech interaction (like gpt-4o)
\end{itemize}
We believe that with the continued scaling of data in the pre-training stage,
a sufficiently strong foundation model can achieve the above three points with only a small amount of high-quality data.

In addition, we will explore deploying TouchTTS on edge devices (such as the D-Robotics Developer Kit X3/X5 (RDK X3/X5) \footnote{https://developer.d-robotics.cc}) to achieve a fully offline speech synthesis system in order to protect user privacy.

In the more distant future, we will also explore voice synthesis guided by facial information as well as style control.